\documentclass[pre,aps,preprint]{revtex4}
\usepackage{graphicx}
\usepackage{amsfonts}
\usepackage{amssymb}
\usepackage{amsmath}
\usepackage{textcomp}

\usepackage{psfrag}

\usepackage{mdwlist}

\usepackage{subfigure}
\usepackage{chngpage}

\DeclareMathOperator{\tr}{tr}

\begin{document}

\title{ An Efficient Linear Programming Algorithm to Generate the Densest Lattice Sphere Packings }

\author{\'{E}tienne Marcotte$^{1}$ and Salvatore Torquato$^{1,2,3,4}$}
\email{torquato@princeton.edu}
\affiliation{$^{1}$Department of Physics,
$^{2}$Department of Chemistry, 
$^{3}$ Program in Applied and Computational Mathematics, $^{4}$ Princeton Institute of the Science and Technology of Materials, 
Princeton University, Princeton, New Jersey 08544, USA}

\begin{abstract}
Finding the densest sphere packing in $d$-dimensional Euclidean
space $\mathbb{R}^d$ is an outstanding fundamental problem with
relevance in many fields, including the ground states of molecular
systems, colloidal crystal structures, coding theory, discrete geometry,
number theory, and biological systems.
Numerically generating the densest sphere packings becomes very challenging in
high dimensions due to an exponentially increasing number of possible
sphere contacts and sphere configurations, even for the restricted
problem of finding the densest lattice sphere packings.
In this paper, we apply the Torquato-Jiao packing algorithm, which
is a method based on solving a sequence of linear programs,
to robustly reproduce the densest known
lattice sphere packings for dimensions 2 through 19.
We show that the TJ algorithm is appreciably more efficient at solving these
problems than previously published methods.
Indeed, in some dimensions, the former procedure can be as much as
three orders of magnitude faster at finding the optimal solutions than
earlier ones. We also study the suboptimal local density-maxima
solutions (inherent structures or ``extreme'' lattices) to gain insight about
the nature of the topography of the ``density'' landscape.
\end{abstract}

\maketitle

\section{Introduction}

There has been great interest in understanding the packings of hard (\emph{i.e.}
nonoverlapping) particles because they serve as useful models for a
variety of many-particle systems arising in the physical and biological
systems, such as liquids
\cite{Hansen1986,Chaikin2000}, glasses
\cite{Zallen1983,Torquato2000,Parisi2010}, crystals 
\cite{Salsburg1962,Torquato2008,Aste2008}, granular media
\cite{Mehta1994,Edwards1999,Torquato2001,Torquato2010b},
and living cells \cite{Torquato2002}.
One outstanding problem is to find the densest packing of identical spheres
in $d$-dimensional Euclidean space $\mathbb{R}^d$.
This seemingly simple problem has proved
to be a challenge for all but the most simple systems; it was not
until 2005 that a proof was successfully presented to confirm the centuries-old
Kepler conjecture \cite{Hales2005}, which states that the densest packing
of spheres in three dimensions is the face-centered cubic lattice.
For $d \ge 4$, there are no proofs for the densest sphere packings,
although for $d=8$ and $d=24$ they are almost surely the $E_8$ and
Leech lattices, respectively \cite{Cohn2009}. Interestingly, these two
lattices have also been used to construct 10- and
26-dimensional string theories, respectively \cite{Gross1985,Chapline1985}.

In recent years, high-dimensional dense sphere packings
have attracted the attention of physicists because of the insights they
offer about condensed-phase systems in lower dimensions
\cite{Andreanov2012,Kallus2010,Torquato2010b,Parisi2010,Torquato2006}.
It is noteworthy that the general problem of
finding the densest sphere packings in $\mathbb{R}^d$ (and other spaces)
is directly relevant to making data transmission over communication channels
resistant to noise \cite{Shannon1948,Conway1999} and of intense
interest in discrete geometry and number theory \cite{Conway1999,Sarnak2006}.
The densest sphere packing problem is also deeply linked to the covering,
quantizer, number variance, and kissing number problems,
with which it shares the best known solutions in a variety of
dimensions \cite{Conway1999,Torquato2010c,Cohn2011}.
Clever analytical methods have been used
to discover dense packings in high dimensions (\emph{i.e.}, $d \ge 4$)
but this approach becomes less efficient as $d$ increases, especially because
lessons learned in lower dimensions cannot be used to construct
packings in higher dimensions \cite{Conway1999,SloaneWebsite}.

Numerical methods have only
recently emerged to discover the densest packings in high-dimensional
spaces.
One such method devised by Kallus, Elser, and Gravel \cite{Kallus2010},
is based on the ``divide and concur''
framework in which a dense arrangement of overlapping spheres is gradually
relaxed until none of the spheres overlap.
Another method formulated by Andreanov and Scardicchio \cite{Andreanov2012}
takes advantage of the fact that all densest lattice packings
are also \emph{perfect lattices} (defined precisely
in Sec.~\ref{sec:results}), which are finite in number
\cite{Conway1999}. The densest lattice packings can therefore be obtained
by randomly exploring the space of perfect lattices.
The efficiency of both algorithms plummets as $d$ grows
larger, preventing them from being effectively used in
very high dimensions \cite{footnote:aboutAndreanov}.

In the past twenty years, the Lubachevsky-Stillinger (LS)
algorithm \cite{Lubachevsky1990} has served as a standard for generating
dense packing of various shaped hard particles
in two and three dimensions \cite{Donev2005,Jiao2008,Jiao2009}.
However, since the LS algorithm is based on a particle-growth molecular
dynamics simulation, it is extremely computationally costly to use it
to generate jammed dense packings with high
numerical accuracy, especially as $d$ grows beyond three dimensions.
A recent improvement on the LS algorithm is the Torquato-Jiao (TJ)
algorithm \cite{Torquato2010a}, which replaces the molecular dynamics with
an optimization problem that is solved using sequential linear programming.
In particular, the density $\phi$ of a sphere packing 
(fraction of space covered by the spheres) within an adaptive
fundamental cell subject to periodic boundary conditions is maximized.
The design variables are the sphere positions (subject to nonoverlap), and
the shape and size of the fundamental cell. The linear programming solution
of this optimization problem becomes exact as the packing approaches the
\emph{jamming point} \cite{Torquato2010b}.
The TJ algorithm has been found to be a very powerful packing protocol
to generate both maximally-dense packings (global maxima) 
and disordered jammed packings (local maxima)
with a large number of identical spheres (per fundamental cell)
across space dimensions \cite{Torquato2010b} as well as maximally
dense binary sphere packings \cite{Hopkins2011,Hopkins2012}.

In this paper, we specialize the TJ algorithm to the restricted problem of
finding the densest \emph{lattice} sphere packings in high dimensions.
In a lattice packing, there is only one sphere per fundamental cell
\cite{footnote:lattice}.
Even this limited problem for $d \ge 4$ brings considerable challenges;
its solution has been proven only for $d \le 8$ \cite{MartinetBook}
and $d = 24$ \cite{Cohn2009}, and it is closely related to the 
shortest-vector problem, which is of NP-hard complexity \cite{Ajtai1996}.
Additionally, most of the densest known sphere packings for $d \le 48$
are lattice packings \cite{Conway1999,SloaneWebsite}.
Tackling the lattice problem is thus a necessary first step prior
to attempting to solve the much more complicated general problem of
finding the densest periodic packings.
A \emph{periodic} packing of congruent particles is obtained
by placing a fixed configuration of $N$ particles where $N > 1͒$ with 
in one fundamental cell of a lattice, which is then periodically replicated
without overlaps.

The outline of the rest of the paper is as follows:
Sec.~\ref{sec:method} describes the implementation of the
TJ algorithm for the special case of lattice sphere
packings. In Sec.~\ref{sec:parameters} we motivate the choices that
we make for the initial conditions and relevant parameters in order
the various problems across dimensions.
In Sec.~\ref{sec:results}, we apply the TJ algorithm for $2 \le d \le 19$,
and show that it is able
to rapidly and reliably discover the densest known lattice packings without
\emph{a priori} knowledge of their existence.
The TJ algorithm is found to be appreciably faster
than previously published algorithms \cite{Kallus2010,Andreanov2012}.
We also demonstrate that the suboptimal-lattice solutions (i.e., the local
maxima ``inherent structures'') are particularly interesting
because they reveal features of the ``density'' landscape.
In Sec.~\ref{sec:conclusion}, we close with some concluding remarks and
a discussion about possible improvements and other applications of the TJ
algorithm.

\section{Application of the TJ algorithm to finding the densest lattice sphere packings\label{sec:method}}

The basic principle behind the TJ algorithm \cite{Torquato2010a}
resides in the fact that finding the densest sphere packing
can be posed as an optimization
problem with a large number of nonlinear constraints
(such as nonoverlap conditions between pairs of particles) 
which can be solved by solving
a series of linear approximations of the original problem.
Its solution eventually converges toward a local or global optimum.
While global optimality cannot be guaranteed, it has been shown that the TJ
algorithm frequently reaches the globally densest packings
\cite{Torquato2010a}. The TJ algorithm was formulated for the general
problem of finding dense periodic sphere packings.
Here we describe its implementation for the special case of determining
the densest lattice sphere packings, which reduces the problem to optimizing
the shape and size of the fundamental cell, since no sphere
translations are involved.
It is interesting to note that the TJ algorithm can be viewed as
a hard-core analog of a gradient descent in the space of lattices
for energy minimizations for systems of particles interacting with soft
potentials as described by Cohn, Kumar, and Sch\"{u}rmann \cite{Cohn2009b}.

Before explaining the numerical details of the TJ algorithm, we need to
define some mathematical quantities. A $d$-dimensional lattice $\Lambda$ is
composed of all vectors that are integer linear combinations of a set of $d$
basis vectors $\mathbf{m}_1$, ..., $\mathbf{m}_d$,
\begin{equation}
\mathbf{P} = n_1 \mathbf{m}_1 + n_2 \mathbf{m}_2 + \cdots + n_d \mathbf{m}_d,
\end{equation}
where $n_j$ are the integers ($j=1,2,\ldots,d$) and we denote by
$\mathbf{n}$ the corresponding column vector with such components.
Using the generator matrix $\mathbf{M}_\Lambda$, whose columns are the basis 
vectors, allows us to explicitly write the lattice set:
\begin{equation}
\Lambda = \left\{ \mathbf{M}_\Lambda \mathbf{n} : \mathbf{n} \in \mathbb{Z}^d \right\} .
\end{equation}
One useful property of $\mathbf{M}_\Lambda$ is that its determinant is equal
(up to a sign) to the volume of the lattice fundamental cell.
We can then write the lattice packing density $\phi$ as the ratio of the
volume occupied by spheres of diameter $D$ to the volume of the
fundamental cell:
\begin{equation}
\phi(\Lambda) = \frac{v(D/2)}{\left| \det \mathbf{M}_\Lambda \right|},
\end{equation}
where
\begin{equation}
v(R)= \frac{\pi^{d/2} R^d}{\Gamma(1+d/2)} 
\end{equation}
is the $d$-dimensional volume of a sphere of radius $R$ and
$\Gamma(n)$ is the Euler gamma function.

The problem of finding the densest lattice packing of spheres
in $d$ dimensions can be expressed as:
\emph{Find the $d\times d$ generator matrix $\mathbf{M}_\Lambda$ with minimal determinant
$|\det \mathbf{M}_\Lambda|$, under the constraint that all non-zero lattice vectors
$\mathbf{M}_\Lambda \mathbf{n}$, $\mathbf{n} \in \mathbb{Z}^d \setminus \{ \mathbf{0} \}$,
are at least as long as $D$.}

For this problem, the Torquato-Jiao algorithm consists of the following
four steps:
\begin{enumerate*}
	\item Randomly create a generator matrix $\mathbf{M}_\Lambda$ according to some
	stochastic process.
	\item For a given \emph{influence sphere radius} $R_I > D$, 
	find all of the non-zero lattice vectors it contains, \emph{i.e.},
	compute $\left\{\mathbf{v} = \mathbf{M}_\Lambda \mathbf{n} : \mathbf{n} \in \mathbb{Z}^d \setminus \{ \mathbf{0} \} \wedge |\mathbf{v}| \le R_I \right\}$.	\label{enum:find_vectors}
	\item Solve a linearized version of a problem, for which the objective
	is to maximize $\phi$
	(equivalent to minimizing $|\det \mathbf{M}_\Lambda|$)
	and the constraints are that none of the vectors calculated in step
	\ref{enum:find_vectors} become shorter than $D$.
	\item Consider whether the algorithm has converged to a lattice that
	is a stable maximum in $\phi$ (either the densest lattice packing
	or a local maximum \emph{inherent structure}
	\cite{footnote:inherent_structure}).
	If it is the former, repeat the procedure
	starting from step \ref{enum:find_vectors}. If it is the latter, the
	solution has converged to a local or global optimum
	and the procedure is terminated.
\end{enumerate*}

In what follows, we provide a more detailed explanation of these four steps.

\subsection{Initialization}\label{sec:init}

There are many possible methods to initialize the generator matrix $\mathbf{M}_\Lambda$. 
Any candidate procedure must both satisfy the minimal length constraint
and adequately sample the space of all lattices. The former is trivially
satisfied by rescaling the matrix if the minimal length constraint is violated.
In order to satisfy the latter condition, we mainly
use Gaussian initial lattices, in which each
coefficient of their generator matrix $\mathbf{M}_\Lambda$ is an independent normal variable
$N(0, \sigma^2)$ with a variance $\sigma^2$. These matrices have the property
that each of their lattice vectors (columns of $\mathbf{M}_\Lambda$) have independent
orientations with no given preference for any particular direction.
To compare this against a different initialization method, we also consider
initial lattices
for which $\mathbf{M}_\Lambda$ is the sum of the generator matrix of a specific lattice packing
(such as the $d$-dimensional checkerboard lattice $D_d$ or
the hypercubic lattice $Z_d$, see Appendix \ref{app:lattice} for the
definitions of these lattices) 
and one of a Gaussian initial lattice.

\subsection{Finding short vectors}

Finding all of the vectors for an arbitrary lattice that are within a small
given radius $R_I$ from the origin is a complex problem in high dimensions.
Indeed, the problem of finding the \emph{shortest lattice vector} for a given
lattice $\Lambda$ grows superexponentially
with $d$ and is in the class of NP-hard (nondeterministic polynomial-time hard)
problems \cite{Ajtai1996}. One efficient method
to solve this problem can be found in Ref.~\onlinecite{Fincke1985}.
The influence sphere radius $R_I$ can be any value larger than the sphere
diameter $D$, and may vary from one iteration to the next.
It is found that the algorithm is largely insensitive to the value chosen
for $R_I$, which is to be contrasted to the results for periodic packings,
where larger $R_I$ values favor the densest packings over inherent structures
\cite{Torquato2010a}.
Since the computational cost of this and the following steps quickly
increases with $R_I$, we opt to use the nearly minimal value $R_I = 1.1 D$.

\subsection{Solving the linearized problem}

The only linearized problem variables in the case of the implementation
of the TJ algorithm in the case of a lattice packing
are the coefficients of the $d\times d$
symmetric strain tensor $\boldsymbol{\varepsilon}$ \cite{footnote:symmetric}.
The modified generator matrix is then
\begin{equation}
\mathbf{M}_\Lambda \rightarrow \mathbf{M}_\Lambda + \boldsymbol{\varepsilon} \mathbf{M}_\Lambda.
\end{equation}
The constraint that a vector originally at position $\mathbf{v} = \mathbf{M}_\Lambda \mathbf{n}$
remains at least as large as $D$ can then be written as
\begin{eqnarray}
\mathbf{n}^\top \mathbf{M}_\Lambda^\top \mathbf{M}_\Lambda \mathbf{n} + 2 \mathbf{n}^\top \mathbf{M}_\Lambda^\top \boldsymbol{\varepsilon} \mathbf{M}_\Lambda \mathbf{n} + \mathbf{n}^\top \mathbf{M}_\Lambda^\top \boldsymbol{\varepsilon}^\top \boldsymbol{\varepsilon} \mathbf{M}_\Lambda \mathbf{n} & \ge & D^2, \nonumber \\
\mathbf{v}^\top \mathbf{v} + 2 \mathbf{v}^\top \boldsymbol{\varepsilon} \mathbf{v} + \mathbf{v}^\top \boldsymbol{\varepsilon}^\top \boldsymbol{\varepsilon} \mathbf{v} & \ge & D^2. \label{eqn:raw_vector_constraints}
\end{eqnarray}
This constraint is linearized by dropping the term that is quadratic in
$\boldsymbol{\varepsilon}$:
\begin{equation}
2 \mathbf{v}^\top \boldsymbol{\varepsilon} \mathbf{v} \ge D^2 - \mathbf{v}^\top \mathbf{v}. \label{eqn:vector_constraints}
\end{equation}
It should be noted that the term
($\mathbf{v}^\top \boldsymbol{\varepsilon}^\top \boldsymbol{\varepsilon} \mathbf{v}$) that has been dropped is non-negative,
which means that every set of variables that satisfies inequality
(\ref{eqn:vector_constraints}) also satisfies inequality
(\ref{eqn:raw_vector_constraints}).
This is different from the equivalent constraints for periodic packings,
for which the quadratic term may be negative due to the interaction between
the lattice deformation and the particle displacements. This avoids the
necessity of either adding a constant term to the constraint or rescaling
the system if spheres are found to overlap, which is the case for the
general periodic packing problem \cite{Torquato2010a}.

Additionally, extra constraints must be added to prevent vectors that could be
outside the influence sphere from becoming shorter than $D$:
\begin{eqnarray}
2 \mathbf{v}^\top \boldsymbol{\varepsilon} \mathbf{v} & \ge & D^2 - R_I^2 \nonumber \\
\frac{\mathbf{v}^\top \boldsymbol{\varepsilon} \mathbf{v}}{\mathbf{v}^\top \mathbf{v}} & \ge & \frac{D^2 / R_I^2 - 1}{2} \equiv -\lambda, \label{eqn:lambda_constraints}
\end{eqnarray}
where the length of the vector has been chosen as its smallest possible
value ($R_I$). A simple yet robust method to ensure that inequality
(\ref{eqn:lambda_constraints}) is satisfied for all vectors outside
of the influence sphere is to bound the lowest eigenvalue
of $\boldsymbol{\varepsilon}$ from below by $-\lambda$.
There are multiple ways to write linear
constraints on $\boldsymbol{\varepsilon}$ such that its eigenvalues are all larger than
$-\lambda$. One such way is given by
\begin{eqnarray}
 -\frac{\lambda}{2} \le \mbox{Diagonal element of $\boldsymbol{\varepsilon}$} < \infty, \label{eqn:diag_constraints} \\
 -\frac{\lambda}{2(d - 1)} \le \mbox{Off-diagonal element of $\boldsymbol{\varepsilon}$} \le \frac{\lambda}{2(d - 1)}. \label{eqn:nondiag_constraints}
\end{eqnarray}
Finally, the determinant of the modified generator matrix (assuming that
$\det \mathbf{M}_\Lambda > 0$) is
\begin{equation}
\det \mathbf{M}_\Lambda \det \left(\mathbf{I} + \boldsymbol{\varepsilon} \right) = \det \mathbf{M}_\Lambda \left(1 + \tr \boldsymbol{\varepsilon} + O(\boldsymbol{\varepsilon}^2)\right),
\end{equation}
where $\mathbf{I}$ is the $d$-dimensional identity matrix.
The linearized density $\phi$ is thus
\begin{equation}
\phi \simeq \phi_0 \left[1 - \tr \boldsymbol{\varepsilon} \right], \label{eqn:objective}
\end{equation}
where $\phi_0$ is the density for the initial generator matrix $\mathbf{M}_\Lambda$ and we
used the fact that the density is inversely proportional to the fundamental
cell volume.
We can see from the above relation that maximizing the lattice density is
equivalent to minimizing the trace of the
strain tensor $\boldsymbol{\varepsilon}$.
Unlike the linearized constraints 
(\ref{eqn:vector_constraints}), (\ref{eqn:diag_constraints})
and (\ref{eqn:nondiag_constraints}), which are conservative in that as long as
they are satisfied the nonlinearized constraints will always be satisfied,
the objective function (\ref{eqn:objective}) may have the wrong sign due
to the nonlinear term having an unknown sign. In the situation where
the updated lattice has a larger determinant than the original matrix,
we halve $\boldsymbol{\varepsilon}$ (multiple times if necessary) to ensure
a lower updated determinant. This prevents the algorithm from oscillating
between multiple lattices and forces it to eventually converge.

\subsection{Convergence criterion}

The algorithm is considered to have converged if the sum of the squared
coefficients of $\boldsymbol{\varepsilon}$ is below a small threshold value
($10^{-12}$ for this paper). This is
numerically equivalent to saying that all lattices in the neighborhood of
the current lattice are less dense. This resulting lattice
is therefore a local density maximum (``inherent structure''
or ``extreme'' lattice, as elaborated in Sec.~\ref{sec:extra_results}).
Such a lattice is also \emph{strictly jammed},
since any possible deformation requires an increase in the volume of its
fundamental cell \cite{Torquato2001,Torquato2003,footnote:strict_jamming}.

\section{Study of parameters and initial conditions\label{sec:parameters}}

The ability of TJ algorithm to discover the densest lattice packings 
can potentially be affected by the influence sphere
radius $R_I$, the lowest eigenvalue of the strain matrix $\lambda$,
and by the choice of the initial lattice.
This section is dedicated to the study of their impact on
the algorithm and to explain our choices for them in the 
following sections.

The TJ algorithm is deterministic \cite{footnote:deterministic}, and therefore
the initial lattice fully controls the resulting final lattice
for given parameters $R_I$ and $\lambda$.
For example, employing initial lattices that are very close to the known
densest lattice, not surprisingly, results in a very high success
rate in obtaining that lattice. On the flip side, it would
almost certainly never be able to discover a hypothetical denser lattice.
It would therefore be misguided to use configurations that are near
the known densest
lattice as the initial conditions. However, allowing initial lattices that
are very bad packers could result in a low success rate or a large convergence
time for success. Thus, good choices
for initial lattices involve a delicate balance between their diversity
and an ability to relax quickly to dense lattices.

\begin{table}[htp]
\begin{adjustwidth}{-1in}{-1in}
	\centering
	\caption{Frequency at which the densest known lattice packing in 13 dimensions, the $K_{13}$ lattice \cite{Conway1999,SloaneWebsite}, is obtained for various parameters using the TJ algorithm. For all sets of influence sphere radii and initial conditions, 10000 lattice packings have been generated, excepted for $R_I = 2.0$ where only 3000 packings were generated. The calculations were performed on a single thread on a 2.40 GHz processor using the Gurobi linear programming library \cite{gurobi}. Since the run time strongly depends on the computer running the program and how well the code is optimized, it should only be used as a rough indication of the program efficiency.}
	\begin{tabular}{|c|c|c|c|}
		\hline
		Sphere of influence radius & Initial conditions & Success rate (\%) & Average time per trial (sec) \\
		\hline
		$R_I = 1.1 D$ & Gaussian & 8.61 & 5.0 \\
		$R_I = 1.1 D$ & $D_{13}$ + noise & 8.21 & 5.5 \\
		$R_I = 1.1 D$ & $\mathbb{Z}^{13}$ + noise & 8.58 & 5.2 \\
		$R_I = 1.1 D$ & Invariant distribution & 8.08 & 29.2 \\
		$R_I = 1.02 D$ & Gaussian & 8.53 & 12.0 \\
		$R_I = 1.5 D$ & Gaussian & 7.61 & 69.9 \\
		$R_I = 2.0 D$ & Gaussian & 6.87 & 1938.5 \\
		variable $R_I$, $\sim 200$ constraints & Gaussian & 7.97 & 6.3 \\
		variable $R_I$, $\sim 2000$ constraints & Gaussian & 7.95 & 17.6 \\
		variable $R_I$, $\sim 2000$ constraints, reduced $\lambda$ & Gaussian & 8.58 & 108.7 \\
		\hline
	\end{tabular}
	\label{tab:varied_params}
\end{adjustwidth}
\end{table}

Table \ref{tab:varied_params} shows numerical results in 13 dimensions.
The initial lattices are taken from four different distributions,
using six different influence sphere radii.
The TJ algorithm typically succeeds at generating the densest known
lattice packing with a high probability. However, it has a relatively lower
success rates for the cases $d=13$ and $d \ge 17$.
We thus purposely choose the 13-dimensional case
to probe the best choices for the initial conditions and algorithmic
parameters because of its abnormally low success rate
in comparison to cases $d \le 16$.
Its low success rate results in better sensitivity to algorithm parameters
compared with dimensions that have naturally higher success rates.
Similar parameter dependence has been observed for other dimensions.

The \emph{Gaussian} initial condition, as previously explained in Sec.~\ref{sec:init},
selects each coefficient of $\mathbf{M}_\Lambda$ from independent normal
distributions with variances $\sigma^2 = D^2$. The initial conditions
referred to as
$D_{d}$ + noise and $\mathbb{Z}^{d}$ + noise starts with the
generator matrices for the checkerboard $D_d$ and hypercubic $\mathbb{Z}^d$
lattices (these lattices are defined in Appendix \ref{app:lattice}),
respectively, with nearest-neighbor distance equal to $D$ plus some noise.
Specifically, we add normal noise
to each coefficient of $\mathbf{M}_\Lambda$ with a variance
$\sigma^2 = D^2 / 100$. The final initial condition type that we attempt to
employ, which we call an \emph{invariant distribution},
generates the lattice from an approximation of the invariant lattice
distribution, using the algorithm described in Ref.~\onlinecite{CohnNotes} with
$p=10007$. For all of these initial conditions, the nearest neighbor distance is
calculated and the lattice is rescaled to avoid any sphere overlap.

As can be seen in Table \ref{tab:varied_params}, the different initial
conditions that we have used result in similar success rates.
We therefore use the Gaussian initial condition to generate the
initial lattices for all subsequent calculations, since it lacks both
the potential bias that the $D_{d}$ + noise and $\mathbb{Z}^{d}$ + noise
initial conditions share, and it does converge much faster than the
invariant distribution.

The main parameter influencing the efficiency of the TJ algorithm is
the influence sphere radius $R_I$, which can either be fixed or vary from
one iteration to the next. A radius that is too large leads to a large
number of extra constraints for the linear program, greatly increasing
its complexity. By contrast, if $R_I$ is too close to $D$, then
the constraints on the shear matrix $\boldsymbol{\varepsilon}$ will be
too restrictive [see Eqs.~(\ref{eqn:lambda_constraints}),
(\ref{eqn:diag_constraints}) and (\ref{eqn:nondiag_constraints})]. This, in
turn, only allows the lattice to deform very slowly, thereby requiring
many iterations before convergence. A compromise between both is to
use a variable $R_I$, such that the number of vectors inside the sphere
of influence stays relatively constant, thus initially allowing a fast
convergence when $\phi$ is small, without needing numerous constraints
when $\phi$ gets close to its maximum. We use the following rough
approximation to select $R_I$:
\begin{equation}
\mbox{Number of constraints} \sim \frac{1}{2} \frac{v(R_I)}{|\det \mathbf{M}_\Lambda|},
\end{equation}
where the factor of one-half comes from the observation that for every
vector $\mathbf{v}$ in a lattice,
there is another one of identical length $-\mathbf{v}$ which does not need
to be explicitly constrained.
A final parameter that can be modified is how much the lattice is allowed
to deform at every iteration. As a test case, we divide the value of
$\lambda$ by 10 to check whether an increased value of $R_I$
provides benefits other than allowing larger strain matrices.

From Table \ref{tab:varied_params}, we can see that increasing $R_I$ does
not increase the success rate (it actually negatively
affects it), while it significantly increases
the run time. Therefore, the following calculations will be done using
a small influence sphere radius of $R_I = 1.1 D$.
We attempted to adjust $R_I$ as a function of dimension $d$ to
improve success rates for large $d$, but this proved to be fruitless.
The radius $R_I$ only weakly impacts the success rate, but its
value has a dramatic influence on the time per trial, which gets
multiplied by 400 when $R_I$ is increased from $1.1 D$ to $2.0 D$.
Therefore, one should decide on a choice of $R_I$ so as to prioritize
a faster execution speed over an increased probability of reaching the
densest lattice packing.

\section{Results\label{sec:results}}

Here we describe the results we obtain by applying the TJ
algorithm to find the densest lattice packings in dimensions 2 through 19.
We compare our results with those obtained in previous
investigations \cite{Kallus2010,Andreanov2012}.
We also provide the frequency of time that the TJ algorithm finds local versus
the densest known global maxima.

\subsection{Finding the densest lattice packings\label{sec:main_results}}

\begin{table}[htp]
	\centering
	\caption{Frequency at which the densest known lattice packing is obtained using the TJ algorithm for $d=2$ through $d=19$ together with the lattices packing fraction $\phi$ and kissing number $Z$. The number of lattice packings generated is 10000 for $d \le 18$ and 100000 for $d = 19$. The influence sphere radius $R_I = 1.1 D$ and the initial lattices are generated using the Gaussian initial condition. See Appendix \ref{app:lattice} for the definitions of the various lattices. The comments in Table \ref{tab:varied_params} concerning computational times also apply here.}
	\begin{tabular}{|c|c|c|c|c|c|c|}
		\hline
		$d$ & \parbox{2.5cm}{Densest\\ lattice packing} & $\phi$ & $Z$ & \parbox{2.5cm}{Success rate\\ (\%)} & \parbox{2.5cm}{Time per trial\\ (sec)} & \parbox{3.5cm}{Time per successful\\ trial (sec)} \\
		\hline
		2 & $A_2$ & 0.9069 & 6 & 100 & $1.7 \times 10^{-5}$ & $1.7 \times 10^{-5}$ \\
		3 & $D_3$ & 0.7405 & 12 & 100 & $8.0 \times 10^{-5}$ & $8.0 \times 10^{-5}$ \\
		4 & $D_4$ & 0.6169 & 24 & 74.31 & $5.6 \times 10^{-4}$ & $7.5 \times 10^{-4}$ \\
		5 & $D_5$ & 0.4653 & 40 & 97.41 & $8.0 \times 10^{-3}$ & $8.2 \times 10^{-3}$ \\
		6 & $E_6$ & 0.3729 & 72 & 89.72 & 0.019 & 0.022 \\
		7 & $E_7$ & 0.2953 & 126 & 91.91 & 0.046 & 0.050 \\
		8 & $E_8$ & 0.2537 & 240 & 84.16 & 0.33 & 0.40 \\
		9 & $\Lambda_9$ & 0.1458 & 272 & 43.82 & 0.21 & 0.49 \\
		10 & $\Lambda_{10}$ & 0.09202 & 336 & 22.74 & 0.49 & 2.1 \\
		11 & $K_{11}$ & 0.06043 & 432 & 19.39 & 1.1 & 5.7 \\
		12 & $K_{12}$ & 0.04945 & 756 & 33.30 & 2.7 & 8.2 \\
		13 & $K_{13}$ & 0.02921 & 918 & 8.61 & 5.0 & 58 \\
		14 & $\Lambda_{14}$ & 0.02162 & 1422 & 20.69 & 10 & 51 \\
		15 & $\Lambda_{15}$ & 0.01686 & 2340 & 23.78 & 16 & 65 \\
		16 & $\Lambda_{16}$ & 0.01471 & 4320 & 22.50 & 51 & 227 \\
		17 & $\Lambda_{17}$ & 0.008811 & 5346 & 1.65 & 55 & $3.4 \times 10^3$ \\
		18 & $\Lambda_{18}$ & 0.005928 & 7398 & 0.10 & 79 & $7.9 \times 10^4$ \\
		19 & $\Lambda_{19}$ & 0.004121 & 10668 & 0.009 & 162 & $1.8 \times 10^6$ \\
		\hline
	\end{tabular}
	\label{tab:succ_rates}
\end{table}

We have applied the TJ algorithm for dimensions $d=2$ through $d=19$,
and found the densest currently known lattice packing for each of them.
The algorithm is robust in that it converges rapidly to the optimal
solutions in most dimensions. Not surprisingly, except for the trivial $d=2$
and $d=3$ cases, it does not reach the optimal solution for all
initial conditions.
Therefore, even though the probabilities of finding the densest packing on the
first attempt was high (greater than 19\% for $d \le 12$ and $14 \le d \le 16$),
we typically needed multiple trials (i.e., different random initial
conditions) to guarantee that the densest
lattice packings were among these. Consequently, the quality of such a
global optimization algorithm is preferably
measured using the time required per successful trial instead of
simply the time per trial or the success rate.
Table \ref{tab:succ_rates} describes the rate at which
the TJ algorithm produced the densest known lattice packings
for dimensions $d=2$ through $d=19$ and the average time required per
successful trial. We determine whether we achieved the
densest known packings primarily by comparing the packing density $\phi$ and
the kissing number $Z$ (the number of spheres that are in contact with any
given sphere) with published data \cite{Conway1999,SloaneWebsite}.
Additionally, we calculate theta series
(the generating functions for the number of vectors with
specific lengths in the lattices \cite{Conway1999})
up through the first few coordination shells.

The time required by the TJ algorithm to generate the densest known
lattice packings is appreciably smaller than the times reported in
Ref.~\onlinecite{Kallus2010}:
approximately 4000 and 25000 seconds per successful packing for
$d=13$ and $d=14$, respectively. The times required by the TJ algorithm of
58 and 51 seconds are orders of magnitude lower, 
indicating a genuine algorithmic improvement that cannot
be attributed to the type of computer employed nor to implementation details.

The authors in Ref.~\onlinecite{Andreanov2012} do not state precise run times
for all dimensions, but report that, after generating more than $10^5$ lattices,
their algorithm is unable to discover the densest known lattices for
$d=14$ through $d=19$. Since generating $10^5$ lattices using their
algorithm takes at least several hours, the TJ algorithm's ability to
successfully generate the densest lattice packings in minutes for $d \le 16$
is a tremendous speed-up improvement.
Using more computing power, the authors in
Ref.~\onlinecite{Andreanov2012} are able to reliably obtain the densest known
lattice for $d \le 17$ using their algorithm \cite{AndreanovCommunication}.
For example, their calculations took four days ($\sim 3 \times 10^5$ seconds)
for $d = 14$, which is three to four orders of magnitude
longer than our own calculations (see Table \ref{tab:succ_rates}).

The fact that the TJ algorithm was unable to find any denser lattice packings
than the densest known lattice packings reinforces the evidence
that these are indeed the densest lattice packings for $d=2$ through $d=19$.
Although this evidence is not as strong for
$d=18$ and $d=19$, due to the rare occurrences of the densest lattice packings,
the evidence is quite strong for $d \le 17$.

One particular aspect of the success rates shown in Table \ref{tab:succ_rates}
is that they do not decrease monotonically with increasing dimension.
Dimensions that are notably difficult are $d=4$ and $d=13$, and neither
case can be explained by lattice packings with unusual properties,
since $d=5$ and $d=12$,
respectively, share similar packings, but not the relatively low success rates.
We will attempt to explain this phenomenon, along with the sharp decrease
in success rates at $d=17$, in the following section.

\subsection{Inherent structures\label{sec:extra_results}}

The TJ algorithm is intrinsically a local density maximization algorithm.
As such, it can, and often does, converge locally to the densest
lattice packing associated with a given initial configuration,
 i.e., an \emph{inherent structure} \cite{Torquato2010a}, that
are not necessarily the global maxima. These local maxima are analogous
to the inherent structures of a continuous potential. 
The study of these inherent structures are of fundamental interest in their
own right because they offer insight about the nature of topography of the ``density''
landscape and understanding the frequency of their occurrence could
potentially lead to improvements on the algorithm.

One interesting property of the density landscape associated with the
lattice packing problem is that all
of its inherent structures are \emph{extreme} lattices, i.e., they
are both \emph{perfect} and \emph{eutactic} \cite{footnote:perfect}.
Only a finite number of distinct extreme lattices exists for any
dimension, which explains how the TJ algorithm is able to always reach the
ground state for $d=2$ and $d=3$, for each of which only a single
extreme lattice exists.
However, as $d$ increases, the number of extreme lattices grows quickly,
possibly exponentially fast. It is thus remarkable
that the TJ algorithm can reliably yield the densest lattice packing
from the large set of possible end states.
This indicates that the ``basin of attraction'' of the ground state
is much larger than the basins of
attraction of the local-maxima inherent structures.
The relatively lower success rates for some dimensions ($d=4$, $d=11$,
$d=13$, and $d \ge 17$) can then be understood as being due to
smaller than usual basins for the corresponding ground states.
The cause of this reduction and whether the symmetry of the inherent
structure is lower than that of
the ground state or some other effect is still unknown and warrants
further investigation.

\begin{table}[htp]
\begin{adjustwidth}{-1in}{-1in}
	\centering
	\caption{Second and third highest-density inherent structures (locally densest lattice packings), including their packing density $\phi$, kissing number $Z$, and success rate from the TJ algorithm. See Table \ref{tab:succ_rates} to compare to the densest lattice packings. The number of lattice packings generated for each dimension is 10000 for $d \le 18$ and 100000 for $d=19$. Multiple lattices with equal density are grouped together and written in ascending kissing number order. See Ref.~\onlinecite{SloaneWebsite} for the definitions of the following lattices: $A_5^{+3}$, $E_6^*$, $P7.3$, $P7.5$, $K_9^2$, Dim11 (named dim11kis422 in the reference), $K_{14}^1$, $K_{14}^2$, $\Lambda_{15}^2$, $K_{15}^1$, $\Lambda_{16}^2$, and $K_{16}^1$. Lattices that were not identified in Ref.~\onlinecite{SloaneWebsite} and found here are denoted as $U_d^n$, where $n$ is used to distinguish different lattices at some fixed dimension $d$.}
	\begin{tabular}{|c|cccc|cccc|}
		\hline
		& \multicolumn{4}{c|}{Second densest} & \multicolumn{4}{c|}{Third densest} \\
		$d$ & Lattice & $\phi$ & $Z$ & Rate (\%) & Lattice & $\phi$ & $Z$ & Rate (\%) \\
		\hline
		2 & --- & --- & --- & --- & --- & --- & --- & --- \\
		3 & --- & --- & --- & --- & --- & --- & --- & --- \\
		4 & $A_4$ & 0.5517 & 20 & 25.69 & --- & --- & --- & --- \\
		5 & $A_5^{+3}$ & 0.4136 & 30 & 1.51 & $A_5$ & 0.3799 & 30 & 1.08 \\
		6 & $E_6^*$ & 0.3315 & 54 & 1.53 & $D_6$ & 0.3230 & 60 & 7.70 \\
		7 & $P7.3$ & 0.2143 & 72 & 0.88 & $P7.5$/$D_7$ & 0.2088 & 72/84 & 1.92/0.11 \\
		8 & $U_8^1$ & 0.1691 & 142 & 0.41 & $U_8^2$ & 0.1530 & 116 & 3.75 \\
		9 & $U_9^1$ & 0.1383 & 258 & 2.60 & $K_9^2$ & 0.1190 & 198 & 14.09 \\
		10 & $U_{10}^1$ & 0.08282 & 294 & 0.42 & $U_{10}^2$ & 0.08231 & 308 & 0.05 \\
		11 & Dim11/$\Lambda_{11}^{\mbox{min}}$/$\Lambda_{11}^{\mbox{max}}$ & 0.05888 & 422/432/438 & 5.32/7.80/0.30 & $U_{11}^1$ & 0.05551 & 408 & 0.81 \\
		12 & $\Lambda_{12}^{\mbox{min}}$/$\Lambda_{12}^{\mbox{mid}}$/$\Lambda_{12}^{\mbox{max}}$ & 0.04173 & 624/632/648 & 9.24/2.96/0.03 & $U_{12}^1$/$U_{12}^2$/$U_{12}^3$ & 0.03732 & 550/560/566 & 1.38/0.18/0.05 \\
		13 & $\Lambda_{13}^{\mbox{min}}$/$\Lambda_{13}^{\mbox{mid}}$/$\Lambda_{13}^{\mbox{max}}$ & 0.02846 & 888/890/906 & 12.17/1.50/0.29 & $U_{13}^1$ & 0.02683 & 828 & 2.97 \\
		14 & $U_{14}^1$ & 0.01934 & 1260 & 0.69 & $K_{14}^2$/$K_{14}^1$ & 0.01922 & 1242/1248 & 2.26/0.38 \\
		15 & $\Lambda_{15}^2$/$U_{15}^1$ & 0.01376 & 1872/1890 & 1.57/0.02 & $K_{15}^1$ & 0.01298 & 1746 & 0.92 \\
		16 & $\Lambda_{16}^2$ & 0.01040 & 2982 & 0.69 & $K_{16}^1$/$U_{16}^1$ & 0.009805 & 2772/2820 & 0.67/0.03 \\
		17 & $U_{17}^1$ & 0.007194 & 4266 & 0.63 & $U_{17}^2$ & 0.006661 & 3942 & 0.09 \\
		18 & $U_{18}^1$ & 0.005134 & 6336 & 0.03 & $U_{18}^2$ & 0.004743 & 5820 & 0.02 \\
		19 & $U_{19}^1$ & 0.003686 & 9480 & 0.012 & $U_{19}^2$ & 0.003475 & 8910 & 0.002 \\
		\hline
	\end{tabular}
	\label{tab:inherent}
\end{adjustwidth}
\end{table}

As seen in Table \ref{tab:inherent}, some inherent structures
are degenerate in the sense that multiple lattices share the same packing
density.
A peculiar property that these degeneracies share is that their appearance
rate is far from constant. For example, it goes from 9.24\% for the
$\Lambda_{12}^{\mbox{min}}$ to a mere 0.03\% for the
$\Lambda_{12}^{\mbox{max}}$. Since both of these are laminated lattices,
why does one occurs more frequently than the other? One possible reason
is that for all these degeneracies but one, the lattices
with smaller kissing number are more likely to be generated.
In the case of $\Lambda_{12}^{\mbox{min}}$ and $\Lambda_{12}^{\mbox{max}}$,
their kissing numbers are respectively 624 and 648.
This is consistent with previous work which has shown that for
packings with many particles per fundamental cell,
the TJ algorithm has a propensity to generate isostatic packings from random
initial conditions, where the number of interparticle contacts is equal to
the number of degrees of freedom of the problem \cite{Torquato2010a}.

\begin{figure}[htp]
	\centering
	\subfigure[]{
		\includegraphics[scale=0.44]{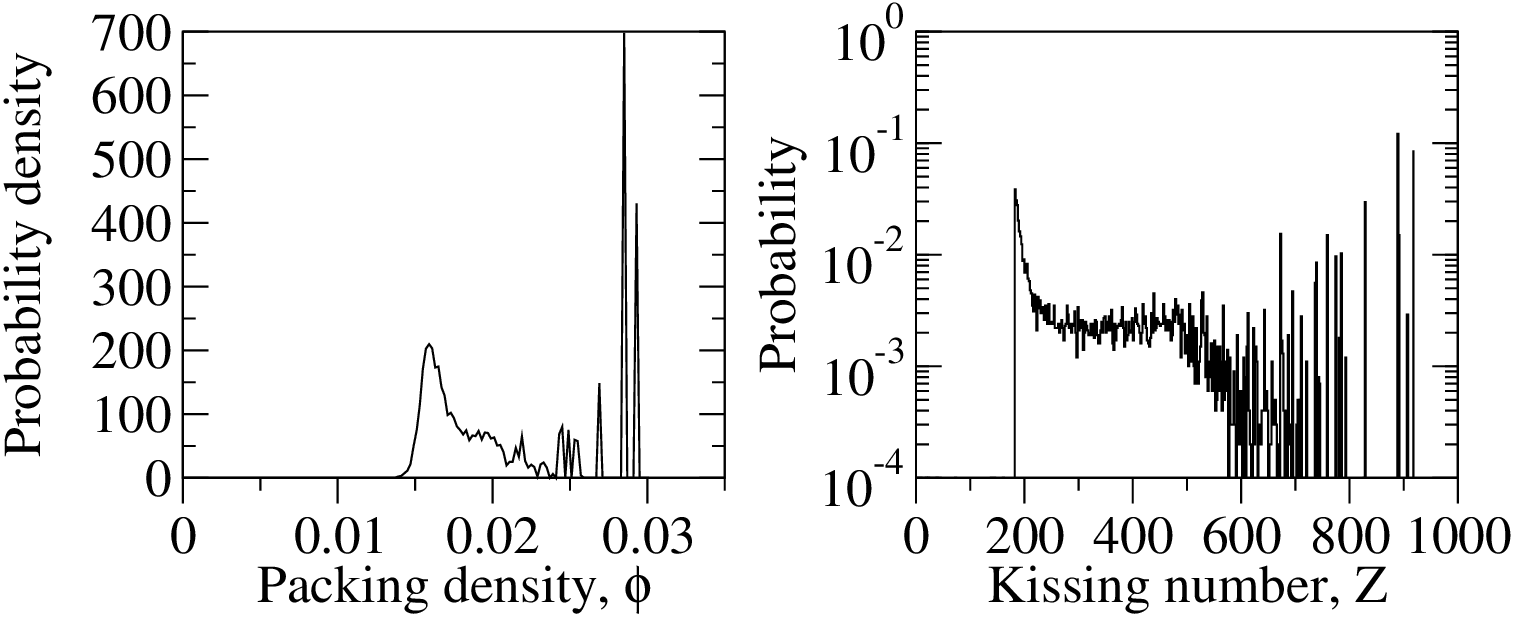}
		\label{fig:phi_density_13}
	}
	\subfigure[]{
		\includegraphics[scale=0.44]{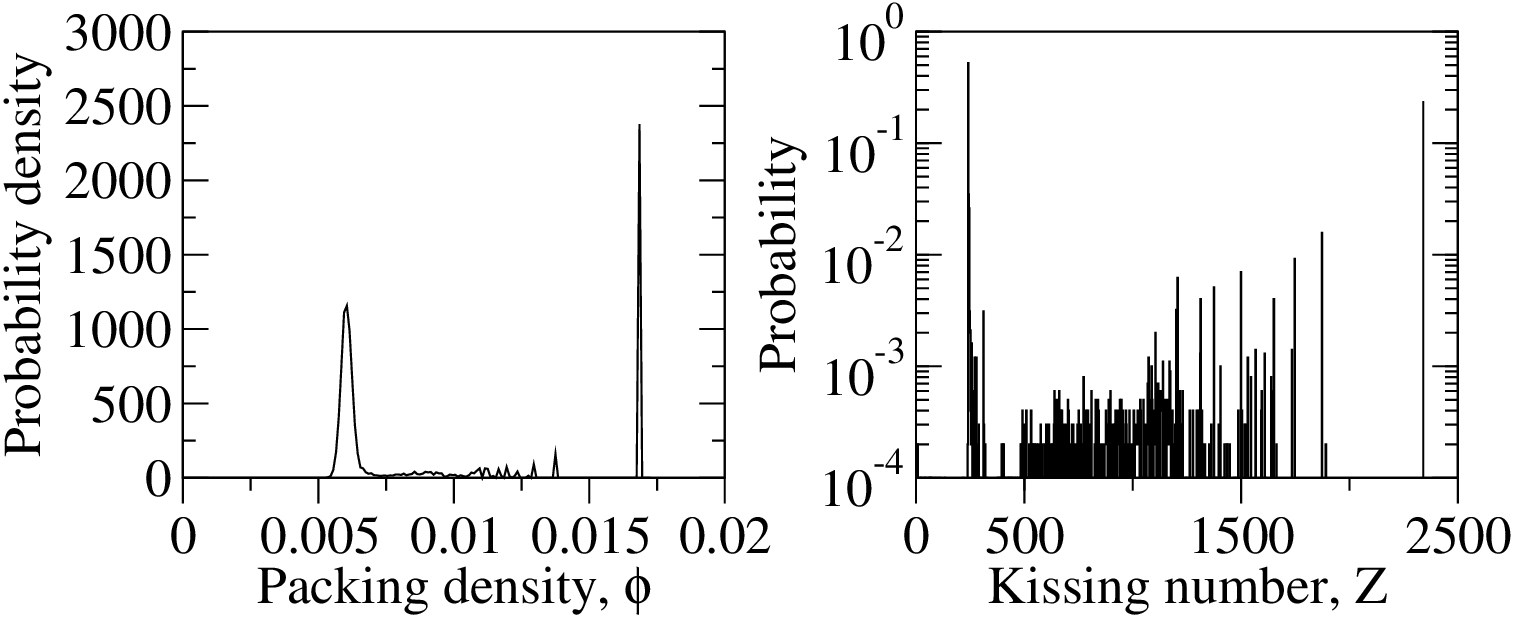}
		\label{fig:phi_density_15}
	}
	\subfigure[]{
		\includegraphics[scale=0.44]{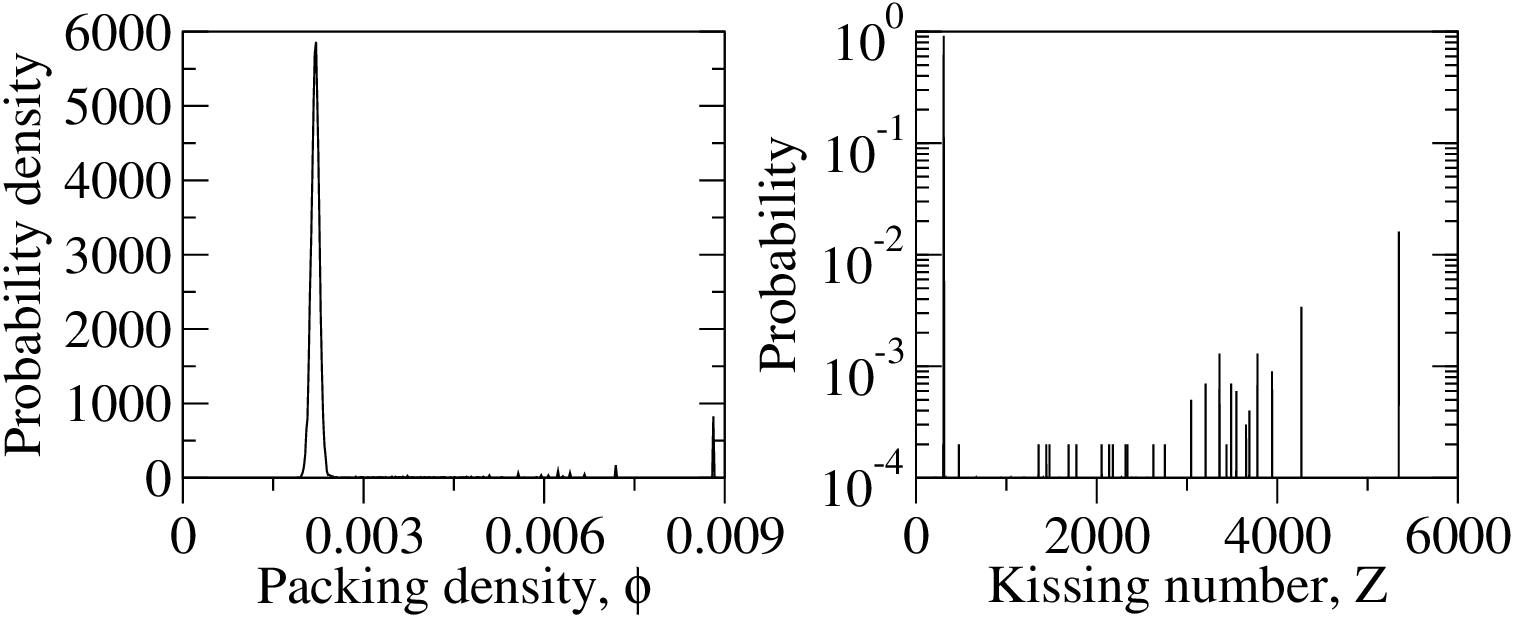}
		\label{fig:phi_density_17}
	}
	\caption{Probability density functions for the packing density $\phi$ (left) and probabilities for the kissing number $z$ (right) of the lattice resulting from the TJ algorithm for \subref{fig:phi_density_13} $d=13$, \subref{fig:phi_density_15} $d=15$, and \subref{fig:phi_density_17} $d=17$. The minimal value of the kissing number $Z_{min}=d(d+1)$ is 182 for $d=13$, 240 for $d=15$, and 106 for $d=17$.}
	\label{fig:phi_density}
\end{figure}

Figure \ref{fig:phi_density} shows that
as the dimensionality increases, the inherent-structure densities tend to
become concentrated around a specific value instead of being spread over a
range of possible densities. This concentration tendency is caused by the
rapid increase in the number of such low-density inherent structures
for large $d$, which eventually overwhelms the algorithmic bias
toward high-density lattices. This explains the dramatic reduction in
success rates in Table~\ref{tab:succ_rates} for $d \ge 17$.
The kissing number has a similar behavior to the packing density,
resulting in the fact that most of the generated lattices for $d \ge 17$ have
an identical low kissing number.
Since these are locally-optimal solutions, a local deformation of the
lattice would either decrease its packing fraction or makes the central sphere
and its neighboring spheres overlap. Therefore, we can define a lower bound
on the kissing number by exploiting the fact that, for a linear program to
have a unique feasible solution,
it requires at least one more active inequality constraint
than the number of degrees of freedom.
Since the problem possesses $d (d+1) / 2$ degrees of freedom
(the number of independent components of $\boldsymbol{\varepsilon}$),
$1 + d (d+1) / 2$ active inequality constraints are required for the problem
to be fully constrained.
One of these constraints comes from the density being at a
local maximum, while each pair of kissing spheres adds a single
constraint. Consequently, the minimum kissing number of a lattice
inherent-structure in $d$ dimensions is $Z_{min} = d (d+1)$.
Referring to Fig.~\ref{fig:phi_density}, we observe that as $d$ increases,
the proportion of generated configurations with a kissing number equal to
$Z_{min}$ increases rapidly relative to all other kissing numbers.
Since the best known lattice packings have high kissing numbers (nearly the
same or equal to highest known kissing numbers), the tendency of the TJ
algorithm tendency to favor lattices with minimal kissing numbers
further explains its low success rates for $d \ge 17$.

\section{Conclusions and Discussion\label{sec:conclusion}}

In this paper, we have shown that the Torquato-Jiao algorithm is able to
quickly find the densest known lattice packings for $d \le 19$. The TJ
algorithm is found to be orders of magnitude faster than the previous
state-of-the-art lattice packing methods \cite{Kallus2010,Andreanov2012}.
This makes the TJ algorithm the fastest current numerical method to generate
the densest lattice packings in high dimensions.

While we limited our present study to $d \le 19$, the TJ algorithm can be
employed to generate dense lattice packings in higher dimensions at
greater computational cost. We expect that dimensions $d=20$ and $d=21$ would
be manageable with more computing resources,
but improvements to the algorithm would be required to study $d \ge 22$.
One possible approach to increase the likelihood of generating a dense
lattice packing for $d \ge 22$ would be to include \emph{ad hoc} methods
in between the TJ-algorithm steps that favor denser packings, such
as thermal equilibration of the system (e.g., via Monte Carlo methods
to solve the ``adaptive shrinking cell'' optimization problem
\cite{Torquato2009,Atkinson2012})
or relaxation under pair potentials known to favor high-density configurations.
Another possibility would be to combine the strengths of the TJ algorithm
with those of other lattice packing methods. The ability of the TJ algorithm
to quickly generate extreme lattices (the inherent structures) could
be used as a starting point for an algorithm that performs an exhaustive
search in the space of perfect lattices \cite{Andreanov2012}. Moreover, its
efficiency in finding locally-densest lattice packings from arbitrary
initial conditions could be used to rapidly obtain such packings starting
from intermediate-density packings generated using other methods
\cite{Kallus2010}.
As $d$ increases from one, the first dimension in which the
densest known packing that is not a Bravais lattice
(periodic packing with a multiple-particle basis) is $d=10$, which has
a basis of 40.
Since the TJ algorithm was successfully used to obtain the
densest known packings for $d \le 6$ with
a large multiple-particle basis (up to a basis
of 729 for $d=6$) \cite{Torquato2010a}, it would be interesting to explore
whether the TJ algorithm could be used to discover currently unknown denser
non-lattice packings in 10 dimensions or higher.

For $d \ge 17$ dimensions, the TJ algorithm mainly produces
lattices that have both a low packing density and a minimal kissing number,
while still being locally densest, revealing a richer and more complex
density landscape than in most dimensions less than 17.
This phenomenon could possibly be exploited to quickly generate
low-density extreme lattices in very high dimensions.
Since these lattices are strictly jammed and have the minimal kissing number
to ensure mechanical stability,
they can be considered to be the lattice analogs to the
maximally random jammed packings (disordered local-maxima inherent structures)
that have been generated using the TJ algorithm with many particles
per fundamental cell \cite{Torquato2010a}.
Such configurations could be generated in much higher dimensions than those
considered in this paper, since the requirement of reaching the ground state
would be removed, and the TJ algorithm is less resource-intensive when
generating suboptimal kissing configurations (through the reduced number of
constraints).

\section*{Acknowledgments}

We thanks Alexei Andreanov, Henry Cohn, Veit Elser, Yoav Kallus,
and Antonello Scarrdicchio for very helpful discussions.
This work was partially supported by the  Materials Research Science
and Engineering
Center Program of the National Science Foundation under Grant
No. DMR- 0820341 and by the Division of Mathematical Sciences at
the National Science Foundation under Award No. DMS-1211087.
This work was partially supported by a grant from the Simons Foundation
(Grant No. 231015 to Salvatore Torquato).
S.T. also thanks the Department of Physics and Astronomy at the University of
Pennsylvania for their hospitality during his stay there.

\appendix
\section{Lattice definitions\label{app:lattice}}

In this appendix, we define some common lattices, following the notation
and nomenclature used in Refs. \onlinecite{Conway1999} and
\onlinecite{SloaneWebsite}.

The \emph{hypercubic} $\mathbb{Z}^d$ lattice is defined by
\begin{equation}
\mathbb{Z}^d=\{(x_1,\ldots,x_d): x_i \in {\mathbb{ Z}}\} \quad \mbox{for}\; d\ge 1
\end{equation}
where $\mathbb{Z}$ is the set of integers ($\ldots -3,-2,-1,0,1,2,3\ldots$)
and $x_1,\ldots,x_d$ denote the components of a lattice vector.
The kissing number of $\mathbb{Z}^d$ is $2d$.
A $d$-dimensional generalization of the face-centered-cubic lattice
is  the \emph{checkerboard} $D_d$ lattice defined by
\begin{equation}
D_d=\{(x_1,\ldots,x_d)\in \mathbb{Z}^d: x_1+ \cdots +x_d ~~\mbox{even}\} \quad \mbox{for}\; d\ge 2.
\end{equation}
Its kissing number is $2d(d-1)$. 
Note that $D_2$ is simply the square lattice $\mathbb{Z}^2$.
Another generalization of the face-centered-cubic lattice is the
\emph{root} lattice $A_d$, which is a subset of points in $\mathbb{Z}^{d+1}$, i.e.,
\begin{equation}
A_d=\{(x_0,x_1,\ldots,x_d)\in \mathbb{Z}^{d+1}: x_0+ x_1+ \cdots +x_d =0\} \quad \mbox{for}\; d\ge 1.
\end{equation}
The kissing number of $A_d$ is $d(d+1)$. In three dimensions, $D_3$ and
$A_3$ are identical, but $D_d$ and $A_d$ are inequivalent for $d\ge 4$.
Another set of root lattices is denoted $E_d$, for $d=6$, $d=7$, and $d=8$.
The root lattice $E_8$ is equal to the union of $D_8$ and the
translation of $D_8$ by
$(\frac{1}{2},\frac{1}{2},\frac{1}{2},\frac{1}{2},\frac{1}{2},\frac{1}{2},
\frac{1}{2},\frac{1}{2})$.
The root lattice $E_7$ is the section of $E_8$ where the sum of the
lattice coefficients is set equal to zero, and the root lattice 
$E_6$ is the section of $E_7$
where the sum of the first and eight coefficients is also set equal to zero.
Alternatively, vectors in $E_8$ perpendicular to any $A_2$-sublattice in $E_8$
also form $E_6$.

The \emph{laminated} lattice $\Lambda_d$ is constructed by stacking layers
of a $(d-1)$-dimensional laminated lattice $\Lambda_{d-1}$ as densely as
possible such that the
shortest vector in $\Lambda_d$ is of equal or longer length than the
shortest vector in $\Lambda_{d-1}$.
This definition does not uniquely define $\Lambda_d$ for all dimensions.
For $d=11$, $d=12$, $d=13$, and $d \ge 25$, there exist multiple laminated
lattices of equal densities, which we distinguish using superscripts.
Many of the previously defined lattices
are also laminated lattices. For example, $\Lambda_1 = \mathbb{Z}^1$,
$\Lambda_2 = A_2$, $\Lambda_3 = D_3$, $\Lambda_4 = D_4$, $\Lambda_5 = D_5$,
$\Lambda_6 = E_6$, $\Lambda_7 = E_7$, and $\Lambda_8 = E_8$.
A particularly interesting laminated lattice is the 24-dimensional
Leech lattice $\Lambda_{24}$.
Finally, the \emph{Coxeter-Todd} lattice $K_{12}$ can be defined
for 18 dimensions:
\begin{equation}
K_{12} = \left\{ (x_{11}, \cdots, x_{16}, x_{21}, \cdots, x_{26}, x_{31}, \cdots, x_{36}) : x_{ij} \in \mathbb{Z} \right\},
\end{equation}
where $x_{ik}$ denotes the components of a lattice vector, 
subject to the following conditions
\begin{eqnarray}
x_{i1} + x_{i2} + x_{i3} = 0 & & i \in \{1, \cdots, 6\}, \\
x_{i1} - x_{j1} \equiv x_{i2} - x_{j2} \equiv x_{i3} - x_{j3} \mod 3 & & i,j \in \{1, \cdots, 6\}, \mbox{and} \\
x_{1k} + x_{2k} + x_{3k} + x_{4k} + x_{5k} + x_{6k} \equiv 0 \mod 3 & & k \in \{1, 2, 3\}.
\end{eqnarray}
This lattice can be generalized to other dimensions in the range $6 \le d \le 18$
by requiring that $K_d$ is the densest section of $K_{d+1}$ which either
contains or is contained in $K_{12}$ and taking $K_{18} = \Lambda_{18}$.


\begin{thebibliography}{26}

\bibitem{Hansen1986}
J. P. Hansen and I. R. McDonald, \emph{Theory of Simple Liquids} (Academic, New York, 1986͒). 

\bibitem{Chaikin2000}
P. M. Chaikin and T. C. Lubensky, \emph{Principles of Condensed Matter Physics} (Cambridge University Press, New York, 2000͒). 

\bibitem{Zallen1983}
R. Zallen, \emph{The Physics of Amorphous Solids} (Wiley, New York, 1983͒). 

\bibitem{Torquato2000}
S. Torquato, T. M. Truskett, and P. G. Debenedetti, Phys. Rev. Lett. 84, 2064 (2000͒). 

\bibitem{Parisi2010}
G. Parisi and F. Zamponi, Rev. Mod. Phys. 82, 789 (2010͒).

\bibitem{Salsburg1962}
Z. W. Salsburg and W. W. Wood, J. Chem. Phys. 37, 798 (1962͒). 

\bibitem{Torquato2008}
S. Torquato and F. H. Stillinger, J. Appl. Phys. 102, 093511 (2007͒); 103, 129902 ͑(2008͒). 

\bibitem{Aste2008}
T. Aste and D. Weaire, \emph{The Pursuit of Perfect Packing} (Taylor \& Francis, New York, 2008͒). 

\bibitem{Mehta1994}
A. Mehta, \emph{Granular Matter: An Interdisciplinary Approach} (Springer-Verlag, New York, 1994͒).

\bibitem{Edwards1999}
S. F. Edwards and D. V. Grinev, Phys. Rev. Lett. 82, 5397 (1999͒); Chem. Eng. Sci. 56, 5451 ͑(2001͒). 

\bibitem{Torquato2001}
S. Torquato and F. H. Stillinger, J. Phys. Chem. B 105, 11849 (2001͒).

\bibitem{Torquato2010b}
S. Torquato and F. H. Stillinger, Rev. Mod. Phys. 82, 2633 (2010͒). 

\bibitem{Torquato2002}
S. Torquato, \emph{Random Heterogeneous Materials: Microstructure and Macroscopic Properties} ͑(Springer, New York, 2002͒).


\bibitem{Hales2005}
T. C. Hales, Ann. Math., \textbf{162}, 1065 (2005).

\bibitem{Cohn2009}
H. Cohn and A. Kumar, Ann. Math., \textbf{170}, 1003 (2009).

\bibitem{Gross1985}
D. J. Gross, J. A. Harvey, E. Martinec, and R. Rohm, Phys. Rev. Lett. \textbf{54}, 502 (1985).

\bibitem{Chapline1985}
G. Chapline, Phys. Lett. B \textbf{158}, 393 (1985).

\bibitem{Torquato2006}
S. Torquato and F. Stillinger, Exp Math. \textbf{15}, 307 (2006); A. Scardicchio, F. H. Stillinger, and S. Torquato, J. Math. Phys. \textbf{49}, 043301 (2008).

\bibitem{Kallus2010}
Y. Kallus, V. Elser, and S. Gravel, Phys. Rev. E \textbf{82}, 056707 (2010).

\bibitem{Andreanov2012}
A. Andreanov and A. Scardicchio, Phys. Rev. E, \textbf{86}, 041117 (2012).

\bibitem{Shannon1948}
C. E. Shannon, Bell Syst. Tech. J. \textbf{27}, 379 ͑(1948͒); \textbf{27}, 623 (1948͒).

\bibitem{Conway1999}
J. H. Conway and N. J. A. Sloane, \emph{Sphere Packings, Lattices, and Groups} (Springer, New York, 1998).

\bibitem{Sarnak2006}
P. Sarnak and A. Str\"{o}mbergsson, Invent. Math. \textbf{165}, 115 (2006).

\bibitem{Torquato2010c}
S. Torquato, Phys. Rev. E \textbf{82}, 056109 (2010).

\bibitem{Cohn2011}
H. Cohn, Y. Jiao, A. Kumar, and S. Torquato, Geom. Topology \textbf{15},
2235 (2011).

\bibitem{SloaneWebsite}
G. Nebe and N. J. A. Sloane, \emph{Catalogue of Lattices}, \\
\texttt{http://www.math.rwth-aachen.de/\textasciitilde{}Gabriele.Nebe/LATTICES/}.

\bibitem{footnote:aboutAndreanov}
It should be noted that the algorithm presented in 
Ref.~\onlinecite{Andreanov2012} was designed to explore general
suboptimal perfect lattice packings (not necessarily the densest
lattice packings) and whether their statistics have implications
for bounds on the maximal density.

\bibitem{Lubachevsky1990}
B. D. Lubachevsky and F. H. Stillinger, J. Stat. Phys. \textbf{60}, 561 (͑1990͒).

\bibitem{Donev2005}
A. Donev, S. Torquato, and F. H. Stillinger, J. Comput. Phys. \textbf{202}, 737 ͑(2005͒); \textbf{202}, 765 ͑(2005͒).

\bibitem{Jiao2008}
Y. Jiao, F. H. Stillinger, and S. Torquato, Phys. Rev. Lett. \textbf{100}, 245504 (2008).

\bibitem{Jiao2009}
Y. Jiao, F. H. Stillinger, and S. Torquato, Phys. Rev. E \textbf{79}, 041309 (2009).


\bibitem{Torquato2010a}
S. Torquato and Y. Jiao, Phys. Rev. E, \textbf{82}, 061302 (2010).

\bibitem{Hopkins2011}
A. B. Hopkins, Y. Jiao, F. H. Stillinger, and S. Torquato, Phys. Rev. Lett. \textbf{107}, 125501 (2011).

\bibitem{Hopkins2012}
A. B. Hopkins, F. H. Stillinger, and S. Torquato, Phys. Rev. E \textbf{85}, 021130 (2012).

\bibitem{footnote:lattice}
In the physical sciences and engineering, a lattice is usually referred as
a Bravais lattice.

\bibitem{Ajtai1996}
M. Ajtai, \emph{Generating Hard Instances of Lattice Problems}, Proc. 28th Annual ACM Symp. Theory of Computing, (1996).

\bibitem{footnote:inherent_structure}
Finding the lattice which maximizes $\phi$ is equivalent to determining the
ground states (global minima) in the ``energy landscape'' in which the
``energy'' is $-\phi$, where the degrees of freedom
are the components of $\mathbf{M}_\Lambda$. Following Torquato
and Jiao \cite{Torquato2010a}, we call the stable local/global
density maxima (or energy minima) \emph{inherent structures}.

\bibitem{Cohn2009b}
H. Cohn, A. Kumar, and A. Sch\"{u}rmann, Physical Review E \textbf{80}, 061116 (2009). 

\bibitem{Fincke1985}
U. Fincke and M. Pohst, Math. Comp., \textbf{44}, 463 (1985).

\bibitem{footnote:symmetric}
An asymmetric shear tensor $\boldsymbol{\varepsilon}$ could have been used,
but the set of inherent-structure solutions would have been unchanged.
Under this more general asymmetric form, in all of the
linearized Eqs. (\ref{eqn:vector_constraints}), (\ref{eqn:diag_constraints}),
(\ref{eqn:nondiag_constraints}), and (\ref{eqn:objective}),
$\boldsymbol{\varepsilon}$ would be replaced by the symmetrized
tensor $(\boldsymbol{\varepsilon} + \boldsymbol{\varepsilon}^\top)/2$.
The only impact from the antisymmetric portion of such an
$\boldsymbol{\varepsilon}$ would be to
add trivial $d(d-1)/2$ rotational degrees of freedom to $\Lambda$, which are
irrelevant as far as the packing density is concerned.

\bibitem{Torquato2003}
S. Torquato, A. Donev, and F. H. Stillinger,  Int. J. Solids Struct. \textbf{40}, 7143 (2003).

\bibitem{footnote:strict_jamming}
The strict jamming of the resulting configuration can only be guaranteed as
long as the packing is restricted to a lattice, i.e.,
one sphere per fundamental cell. For any such strictly jammed
lattice sphere packing, it is possible that
density-preserving or density-increasing deformations exist
on a larger torus involving this structure that would lead to
unjamming motions. This possibility increases with increasing dimension,
since all lattice packings almost surely become ``unsaturated''
(holes exist that can accommodate extra spheres) in sufficiently high
dimensions \cite{Conway1999}.

\bibitem{footnote:deterministic}
It is important to realize that the linear programs that the TJ algorithm
solves often do not have unique solutions.
Therefore, the TJ algorithm is guaranteed to be deterministic
only if the selection of an optimal solution is also guaranteed to be
deterministic.
An example of multiple equivalent solutions occurs
if all of the lattice non-zero vector lengths are larger than
the sphere of influence radius $R_I$,
in which case the only linearized problem constraints will be those
on the shear matrix $\boldsymbol{\varepsilon}$.
Then, as long as they satisfy inequality (\ref{eqn:nondiag_constraints}),
the non-diagonal elements of $\boldsymbol{\varepsilon}$ can
take any value, without any impact on the objective function.

\bibitem{gurobi}
\emph{Gurobi Optimizer}, version 5.0.2, Gurobi Optimization, \texttt{www.gurobi.com}.

\bibitem{CohnNotes}
D. Goldstein and A. Mayor, Forum Math. \textbf{15}, 165 (2003);
H. Cohn and G. Minton, private communication, 2013.

\bibitem{AndreanovCommunication}
A. Andreanov and A. Scardicchio, private communication, 2013.

\bibitem{footnote:perfect}
\emph{Perfect} lattices in $d$ dimensions have the property that any 
$d \times d$ symmetric matrix $S$ can be written as a linear combination of the
lattice shortest vector projectors $\mathbf{v}^{min}_i$, \emph{i.e.}
\begin{equation}
S = \sum_{i=1}^Z \alpha_i \mathbf{v}^{min}_i \mathbf{v}^{min \top}_i, \nonumber
\end{equation}
where $Z$ is the lattice kissing number and $\alpha_i$ are linear
combination coefficients.
All perfect lattices for $1 \le d \le 8$ have been identified.
There are 1, 1, 1, 2, 3, 7, 33, and 10916 perfect lattices
for $d=1$ through $d=8$, respectively \cite{SloaneWebsite}.
\emph{Eutactic} lattices in $d$ dimensions have the property than each of their
shortest vector $\mathbf{v}^{min}_i$ is associated with a positive
\emph{eutactic coefficient} $\beta_i > 0$ such that the norm of any vector
$\mathbf{x}$ can be written as:
\begin{equation}
|\mathbf{x}|^2 = \sum_{i=1}^Z \beta_i \left( \mathbf{v}^{min \top}_i \mathbf{x} \right)^2. \nonumber
\end{equation}
A lattice is a local maximum in density (i.e., an inherent structure)
if and only if it is an \emph{extreme} lattice, which is both perfect and
eutactic \cite{Voronoi1908}.
There are 1, 1, 1, 2, 3, 6, 30, and 2408 extreme lattices for $d=1$ through
$d=8$, respectively \cite{MartinetBook,Riener2006}.

\bibitem{Voronoi1908}
G. Voronoi, J. reine angew. Math. \textbf{133}, 97 (1908).

\bibitem{MartinetBook}
J. Martinet, \emph{Perfect lattices in Euclidean Spaces} (Springer, New York, 2003).

\bibitem{Riener2006}
C. Riener, J. Th. Nombres Bordeaux \textbf{18}, 677 (2006).

\bibitem{Torquato2009}
S. Torquato and Y. Jiao, Phys. Rev. E \textbf{80}, 041104 (2009).

\bibitem{Atkinson2012}
S. Atkinson, Y. Jiao, and S. Torquato, Phys. Rev. E \textbf{86}, 031302 (2012). 

\end{thebibliography}
\end{document}